# Constraints on the Optical Depth to Reionization from Balloon-borne Cosmic Microwave Background Measurements

Josquin Errard[1], Mathieu Remazeilles[2,3], Jonathan Aumont[4], Jacques Delabrouille[5], Daniel Green[6], Shaul Hanany[7],
Brandon S. Hensley[8], and Alan Kogut[9]
[1] Université Paris Cité, CNRS, Astroparticule et Cosmologie, F-75013 Paris, France
[2] Instituto de Fisica de Cantabria (CSIC-UC), Avda. los Castros s/n, E-39005 Santander, Spain
[3] Jodrell Bank Centre for Astrophysics, The University of Manchester, Oxford Road, Manchester, M13 9PL, U.K.
[4] IRAP, Université de Toulouse, CNRS, UPS, CNES 9, Avenue du Colonel Roche, BP 44346, F-31028, Toulouse Cedex 4, France
[5] CNRS-UCB International Research Laboratory, Centre Pierre Binétruy, IRL2007, CPB-IN2P3, Berkeley, CA 94720, USA
[6] Department of Physics, University of California at San Diego, La Jolla, CA 92093, USA
[7] University of Minnesota—Twin Cities, 115 Union St. SE, Minneapolis, MN 55455, USA; hanany@umn.edu
[8] Department of Astrophysical Sciences, Princeton University, Princeton, NJ 08544, USA
[9] NASA Goddard Space Flight Center, Greenbelt, MD 20771, USA
Received 2022 June 5; revised 2022 October 10; accepted 2022 October 10; published 2022 November 22

## Abstract

We assess the uncertainty with which a balloon-borne experiment, nominally called Tau Surveyor ($\tau$S), can measure the optical depth to reionization $\sigma(\tau)$ with given realistic constraints of instrument noise and foreground emissions. Using a $\tau$S fiducial design with six frequency bands between 150 and 380 GHz, with white and uniform map noise of 7 $\mu$K arcmin, achievable with a single midlatitude flight, and including Planck's 30 and 44 GHz data, we assess the error $\sigma(\tau)$ obtained with three foreground models and as a function of sky fraction $f_{sky}$ between 40% and 54%. We carry out the analysis using both parametric and blind foreground separation techniques. We compare the $\sigma(\tau)$ values to those obtained with low-frequency and high-frequency versions of the experiment called $\tau$S-lf and $\tau$S-hf, which have only four and up to eight frequency bands with narrower and wider frequency coverage, respectively. We find that with $\tau$S, the lowest constraint is $\sigma(\tau) = 0.0034$, obtained for one of the foreground models with $f_{sky} = 54\%$. $\sigma(\tau)$ is larger, in some cases by more than a factor of 2, for smaller sky fractions, with $\tau$S-lf, or as a function of foreground model. The $\tau$S-hf configuration does not lead to significantly tighter constraints. The exclusion of the 30 and 44 GHz data, which give information about synchrotron emission, leads to significant $\tau$ misestimates. Decreasing noise by an ambitious factor of 10, while keeping $f_{sky} = 40\%$, gives $\sigma(\tau) = 0.0031$. The combination of $\sigma(\tau) = 0.0034$, baryon acoustic oscillation data from DESI, and future cosmic microwave background B-mode lensing data from the CMB-S3/CMB-S4 experiments could give $\sigma(\sum m_\nu) = 17$ meV.

*Unified Astronomy Thesaurus concepts:* Cosmic microwave background radiation (322); Reionization (1383); Neutrino masses (1102); High altitude balloons (738)

## 1. Introduction

The optical depth to reionization $\tau$ is one of several parameters encoding the evolution of cosmological structure formation. Measurements of $\tau$ inform studies of the physical processes during the epoch of reionization (Cooray et al. 2019) and are a key ingredient in cosmological constraints on the sum of neutrino masses (Dvorkin et al. 2019). The optical depth is directly measurable through observations of the E- and B-mode polarization of the cosmic microwave background (CMB) radiation on large angular scales $\ell \lesssim 20$. At these angular scales, the E-mode level is at least a factor of 50 larger than the B mode, which has not yet been detected, and for the remainder of this paper we focus on E-mode measurements. Planck Collaboration et al. (2020a) report $\tau = 0.054 \pm 0.007$, based on data from the Planck space mission, and a more recent analysis of the same data by Pagano et al. (2020) gives $\sigma(\tau) = 0.006$.

Measurements of solar and atmospheric neutrinos, and results from long-baseline and nuclear reactor data probing neutrino oscillations on various length scales, have established that at least two species of neutrinos have mass (de Salas et al. 2018; Tanabashi et al. 2018), and that at least one is now nonrelativistic. They pointed to two possible mass hierarchies among the three neutrino species and determined the differences between the squares of the masses. The absolute mass scale, however, is still unknown, and the standard model of particle physics has no explanation for the origin of neutrino mass. For a normal mass hierarchy, in which the two neutrinos with the smallest mass difference have the lowest mass, the total mass is constrained to $\sum m_\nu \geqslant 58$ meV, while for an inverted hierarchy, for which the two neutrinos with the smallest mass difference are more massive than the third, the total mass is constrained to $\sum m_\nu \geqslant 110$ meV. Finding the total mass of the neutrinos and determining whether the mass hierarchy is normal or inverted are key to understanding the structure of matter.

Over the next decade, cosmological measurements are poised to provide the strongest constraints on the absolute mass scale, by determining the sum of neutrino masses. Cosmology constrains $\sum m_\nu$ because the mass of the neutrinos affects the clustering of matter in a scale-dependent way (Hu et al. 1998). Gravitational lensing, as revealed either in maps of the CMB or in the correlated distortion of galaxy shapes in deep, wide-field galaxy surveys, gives a sensitive probe to the clustering properties of matter (Abazajian & Dodelson 2003; Kaplinghat et al. 2003). However, the clustering of matter is also sensitive to the density of matter $\omega_m$ and to the initial (primordial) amplitude of fluctuations $A_s$, set by inflation. As a result, the measurement of $\sum m_\nu$ is limited by our knowledge of these two parameters. Forthcoming measurements of baryon acoustic oscillations (BAOs) by DESI and Euclid (DESI Collaboration et al. 2016; Amendola et al. 2018) will soon give sufficiently high-accuracy determinations of $\omega_m$ (Pan & Knox 2015). CMB surveys by experiments at the Simons Observatory (SO) and CMB-S4 (Abazajian et al. 2016; The Simons Observatory Collaboration et al. 2018) will provide high-precision measurements of the combination $A_s e^{-2\tau}$. An additional direct, tight measurement of $\tau$ is necessary to constrain $A_s$ and give a determination stronger than $3\sigma$ on $\sum m_\nu$ (Allison et al. 2015). A measurement exceeding this threshold would impact our understanding of the history of the universe and would place stringent constraints on a wide range of new particles and forces (Green & Meyers 2021).

The combination $A_s e^{-2\tau}$ has been measured by Planck to an accuracy of 0.6%, while Planck's error on $\sigma(\tau)$ is 13% (Planck Collaboration et al. 2020a), giving a comparable error on $A_s$. These constraints, combined with BAO observations by Sloan Digital Sky Survey IV, constrain the sum of neutrino masses to $\sum m_\nu < 0.115$ eV (95% confidence limit; Alam et al. 2021). To make further improvements on constraining $\sum m_\nu$, it is essential to reduce the error on $\tau$. The most promising way of achieving higher-precision measurements of $\tau$ is to map E modes over large portions of the sky to a depth better than Planck's. In the future, it might also be possible to constrain $\tau$ with kinetic Sunyaev–Zeldovich (Alvarez et al. 2021) and with hydrogen 21 cm data (Liu et al. 2016). E-mode measurements over large portions of the sky must contend with foreground sources of emission. For measurements between 30 and 100 GHz, Galactic synchrotron emission dominates, and measurements at higher frequencies are dominated by emission from polarized Galactic dust. At 40 and 220 GHz, and over the cleanest 78% of the sky, E-mode synchrotron and dust emission are expected to be 30 and 40 times stronger than the CMB E mode, respectively, at $\ell = 10$ (Planck Collaboration et al. 2020b).

A definitive measurement of $\tau$ is best done from space, because satellites can observe the entire sky and because their frequency coverage is not limited by access to a few atmospheric windows. Space missions such as CORE, LiteBIRD, and PICO are predicted to provide cosmic variance–limited measurements of $\tau$ with $\sigma(\tau) = 0.002$ (Di Valentino et al. 2018; Hanany et al. 2019; LiteBIRD Collaboration et al. 2022). Such a measurement, when combined with BAO data from DESI or Euclid, will give a $4\sigma$ constraint on the minimum sum $\sum m_\nu = 58$ meV (Di Valentino et al. 2018; Hanany et al. 2019; LiteBIRD Collaboration et al. 2022). Nevertheless, there are several funded suborbital instruments attempting to improve upon Planck's measurement. They are the ground-based CLASS instrument, which has been taking data since June 2016 (Dahal et al. 2020, 2021), the balloon-borne PIPER (Holland et al. 2014), and TAURUS.[10] Watts et al. (2018) have assessed the capability of CLASS to measure $\tau$ and constrain $\sum m_\nu$. They used the instrument's planned four frequency bands between 40 and 220 GHz, with a combined noise level of 8 $\mu$K arcmin after 5 yr, to conclude that if CLASS can measure multipoles $\ell \geqslant 2$, it will achieve $\sigma(\tau) = 0.003$. Pawlyk et al. (2018) described the capabilities of PIPER, which has frequency bands between 200 and 600 GHz and plans to survey 85% of the sky. No quantitative constraints were given for measuring $\tau$.

One important feature of balloon-borne instruments is their access to frequency bands above ∼300 GHz. The atmosphere has significant opacity at these frequencies, and atmospheric turbulence induces elevated noise, both of which make ground-based measurements difficult. In this paper, we are investigating the constraints that a balloon-borne instrument can place on $\tau$ during a single ultralong-duration flight, and we are assessing the value added by having access to frequency bands above 300 GHz. Our main focus is accounting for and subtracting Galactic foregrounds, a process that is commonly called "component separation." For concreteness, we will refer to a fiducial balloon experiment called Tau Surveyor ($\tau$S; Section 2). We construct simulated maps as observed by $\tau$S (Section 3.2), and we use two component separation techniques to extract the underlying CMB signal and to estimate $\tau$ and $\sigma(\tau)$ (Section 4). The maps are constructed using three distinct foreground sky models (Section 3.1), all consistent with current observations by Planck. We then change the $\tau$S focal plane, using a version that only has lower frequencies, below 300 GHz, called $\tau$S-lf. We compare the $\tau$S constraints to those obtained with $\tau$S-lf. In Section 5, we discuss the results and compare them to configurations without the 30 and 44 GHz data, with one-tenth the noise, with the same noise but with two additional higher-frequency bands, and with 80% coverage of the sky. We also translate the derived $\sigma(\tau)$ values to constraints on $\Sigma m_\nu$.

## 2. Experiment Configuration

The experiment configuration for $\tau$S is based on a balloon-borne instrument that was proposed as a NASA Pioneer program payload for an ultralong-duration flight from Wanaka, New Zealand. The instrument had a 4 K cooled, two-mirror cross-Dragone telescope, feeding a focal plane with two-color pixels (Hubmayr et al. 2011; Datta et al. 2016), each with four polarization-sensitive bolometers. The focal plane was planned to be maintained at 0.1 K. Several configurations of the instrument were considered, including entrance apertures of 30 or 40 cm in diameter, and options for the total number of detectors, which depended on readout multiplexing factors and power dissipation. The fiducial configuration that we assume in this paper is the one that had the lowest noise—see Table 1. At 150 GHz, the single detector photon noise term (in $W^2/Hz$) is 72% of the total noise. For technical reasons, and to maintain the noise margin, this was not the proposed configuration. Nevertheless, it is useful to consider the lowest noise configuration here, because it represents the lowest limit achievable with this realistic experiment configuration. Noise was assumed to have a white frequency spectrum.

Table 1 gives the specifications for $\tau$S and $\tau$S-lf. $\tau$S has six frequency bands. Three, between 150 and 220 GHz, are designed to provide the signal-to-noise ratio on the measurement of $\tau$, and three, between 260 and 380 GHz, give information about Galactic dust. We rely on 30 and 44 GHz data from Planck to constrain the Galactic synchrotron emission. The number of detectors at each band is based on filling the diffraction-limited field of view of a realistic optical design. The pixel spacing has been chosen to give the highest

---

[10] https://indico.cmb-s4.org/event/27/contributions/405/attachments/402/891/2021-08-13%20Taurus%20%28S4%20Meeting%29.pdf

**Table 1**
Configurations for the $\tau$S/$\tau$S-lf Focal Planes

| Pixel Type | Frequency Band (GHz) | Beam Size (arcmin) | Detector NET[a] ($\mu$K $\sqrt{s}$) | Number of Detectors | Array NET[a] ($\mu$K $\sqrt{s}$) | Polarization Weight ($\mu$K arcmin) |
|---|---|---|---|---|---|---|
| Low        | 150 | 35 | 64  | 4410 | 0.96 | 9.5 |
| Frequency  | 220 | 24 | 87  | 3234 | 1.5  | 15 |
| Middle     | 180 | 29 | 90  | 1800/3000 | 2.1/1.65 | 21/16 |
| Frequency  | 260 | 20 | 141 | 1800/3000 | 3.3/2.6  | 33/25.5 |
| High       | 310 | 17 | 350 | 2028/0 | 7.8/0  | 77/0 |
| Frequency  | 380 | 14 | 833 | 2028/0 | 18.5/0 | 183/0 |
| Total      |     |    | 42  | 15,300/13,644 | 0.74/0.70 | 7.3/6.9 |

**Note.** For $\tau$S-lf, we replace the 310/380 GHz pixels with 180/260 GHz pixels. Because the diameters of the lower-frequency pixels are larger, a smaller number of pixels is added at the lower frequencies than is eliminated from the higher ones.
[a] Noise equivalent temperature, in CMB thermodynamic units.

average mapping speed for each type of two-color pixel. $\tau$S-lf only has the four lower-frequency bands that $\tau$S has. The space that is made available in the focal plane by the deletion of the 507 pixels with the two higher bands has been filled with 300 pixels at the middle band. Adding an even smaller number of pixels to the 800–1000 pixels at the low-frequency bands would have made a negligible difference to the noise. We chose to replace the high-frequency pixels with lower-frequency ones, rather than just remove them and leave the focal plane area empty, because we are attempting to contrast realistic options. In an actual experiment, all the available focal plane area is used. When simulated maps are analyzed, we use Planck–LFI 30 and 44 GHz data, with 3.5 and 4.0 $\mu$K deg polarization weight white-noise levels, respectively (Planck Collaboration et al. 2020c).

We assume a 40 day flight. Only the night portion of any 24 hr period is allocated for $\tau$ scans, which are assumed to be conducted with 360° rotation of the payload. We simulate 12 hr per night scans for the duration of the flight and find that a maximum of 57% of the sky is scanned. The simulations include scanning elevations between 30° and 60°, and given the highly unpredictable nature of the wind patterns for this flight, we assume a constant drift rate for the payload in longitude throughout the 40 day flight from New Zealand to South America. Daytime scans, conducted mostly in the anti-Sun direction, and used to characterize Galactic dust emission in smaller regions of the sky, are not included in the $\tau$ analysis. The noise is assumed to be uniformly distributed over the scanned area. This is an appropriate simplification for the experiment concept preparation; see Section 5.3.

## 3. Foreground Models and Sky Maps

### 3.1. Foreground Models

While the spatial features of the polarized Galactic emission have been measured by Planck at 353 GHz (Planck Collaboration IV 2020), and by both Planck and the Wilkinson Microwave Anisotropy Probe (WMAP) near 100 GHz (Bennett et al. 2013; Planck Collaboration IV 2020), with resolution far exceeding the resolution required to measure $\tau$, there are significant uncertainties about the spatial dependence of the spectral energy distributions (SEDs) of the emission components. It is this uncertainty that limits our knowledge of the foregrounds, and therefore increases the uncertainty of the CMB signal extracted during the component separation process. We capture the possible variability of the real sky by employing a suite of three simulated skies, all of which are consistent with current data, and have similar spatial features, yet have different SEDs. Each of the skies is motivated by a foreground model, as described below, with analytical parameterization expressed in specific intensity units (MJy/sr).

#### 3.1.1. Model d1s1

Model d1s1 employs the "d1" dust emission model and the "s1" synchrotron emission model from the PySM software (Thorne et al. 2017; Zonca et al. 2021). Based on both raw data and component separation products (Miville-Deschênes et al. 2008; Bennett et al. 2013; Planck Collaboration X 2016), this model can be considered the simplest parametric model consistent with Galactic emission, as measured by WMAP and Planck. However, it does not include complexities, such as the line-of-sight variation of dust spectral parameters, that are known to exist in the true sky (Pelgrims et al. 2021).

In the d1 model, the dust emission in each sky pixel $p$ follows a modified blackbody (MBB) emission law:

$$Q_{\nu,p}^{\mathrm{d1}} = A_{d,p}^{Q}\left(\frac{\nu}{\nu_0}\right)^{\beta_{d,p}} B_\nu(T_{d,p})$$
$$U_{\nu,p}^{\mathrm{d1}} = A_{d,p}^{U}\left(\frac{\nu}{\nu_0}\right)^{\beta_{d,p}} B_\nu(T_{d,p}), \quad (1)$$

where $Q_\nu$ and $U_\nu$ are the Stokes parameters at frequency $\nu$, $A_d^{[Q,U]}$ are frequency-independent dust amplitude parameters, $B_\nu$ is the Planck blackbody function, $T_d$ is the dust temperature, $\beta_d$ parameterizes the frequency dependence of the dust opacity, and $\nu_0 = 353$ GHz is a reference frequency. The d1 model employs maps of $A_d^Q$, $A_d^U$, $T_d$, and $\beta_d$, based on component separation of the microwave sky with Haslam, WMAP, and Planck data (Planck Collaboration X 2016). These templates are smoothed to a resolution (FWHM) of 2.6°, and smaller scales are added as Gaussian random fluctuations.

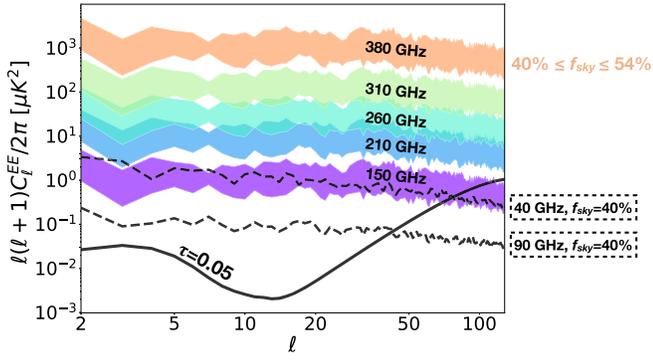

**Figure 1.** E-mode power spectra for model d1s1 between 150 and 380 GHz (colored strips), at 40 and 90 GHz (dashed lines), and for the CMB E-mode power spectrum (black line), with our fiducial value $\tau = 0.05$. Each colored strip brackets the expected foregrounds for $f_{sky} = 40\%$ (bottom of strip) and 54% (top of strip). The dashed lines are for $f_{sky} = 40\%$, and in combination with the strips they show that even near their minima, the foreground levels are larger than the CMB E-mode for $\ell < 40$. At frequencies lower than $\sim 90$ GHz, polarized synchrotron emission dominates.

In the "s1" synchrotron model, the synchrotron emission in each sky pixel scales as a power law in frequency:

$$Q^{s1}_{\nu,p} = A^Q_{s,p} \left(\frac{\nu}{\nu_0}\right)^{\beta_{s,p}+2}$$
$$U^{s1}_{\nu,p} = A^U_{s,p} \left(\frac{\nu}{\nu_0}\right)^{\beta_{s,p}+2}, \quad (2)$$

where $\beta_s$ is the power-law index and $\nu_0 = 23$ GHz is a reference frequency. The s1 model employs maps of $A_s^{[Q,U]}$ from the WMAP 9 yr 23 GHz data (Bennett et al. 2013) and a $\beta_s$ map from "Model 4" of Miville-Deschênes et al. (2008), who derived synchrotron spectral indices using a combination of the 3 yr WMAP 23 GHz data (Hinshaw et al. 2007) and Haslam 408 MHz data (Haslam et al. 1982). The amplitude maps are smoothed to a resolution of 5°, and smaller scales are added as Gaussian random fluctuations. Figure 1 shows the power spectra for d1s1 for different frequency bands and sky masks.

### 3.1.2. Model d7s3

The "d7" dust model uses the same 353 GHz amplitude maps $A_d^{[Q,U]}$ as the "d1" model, including the Gaussian small-scale fluctuations, but it implements a different frequency scaling. Instead of a perfect MBB, the dust emission is modeled as arising from a population of dust grains with different compositions, sizes, and temperatures, as described by Hensley (2015). The frequency dependence is parameterized by a single quantity $\mathcal{U}$, which governs the strength of the radiation field heating the dust. The value of $\mathcal{U}$ in the pixel $p$ is determined from the $\beta_d$ and $T_d$ maps employed by the d1 model via

$$\log_{10}\mathcal{U}_p = (4 + \beta_{d,p}) \log_{10}\left(\frac{T_{d,p}}{\langle T_d \rangle}\right), \quad (3)$$

where $\langle T_d \rangle$ is the mean value of the $T_d$ map. Notable in the d7 model is the presence of ferromagnetic iron inclusions in the grains, which affect the low-frequency ($\lesssim 100$ GHz) polarization spectrum of the dust emission (Draine & Hensley 2013).

Explicitly,

$$Q^{d7}_{\nu,p} = A^Q_{d,p} \frac{f_\nu(\mathcal{U}_p)}{f_{\nu_0}(\mathcal{U}_p)}$$
$$U^{d7}_{\nu,p} = A^U_{d,p} \frac{f_\nu(\mathcal{U}_p)}{f_{\nu_0}(\mathcal{U}_p)}. \quad (4)$$

The function $f_\nu(\mathcal{U})$ that gives the frequency dependence does not have an analytic representation.

The "s3" model of synchrotron emission is identical to the "s1" model in all respects, except for the addition of a curvature term to the frequency scaling law:

$$Q^{s3}_{\nu,p} = A^Q_{s,p} \left(\frac{\nu}{\nu_0}\right)^{\beta_{s,p}+2+C\ln(\nu/\nu_0)}$$
$$U^{s3}_{\nu,p} = A^U_{s,p} \left(\frac{\nu}{\nu_0}\right)^{\beta_{s,p}+2+C\ln(\nu/\nu_0)}, \quad (5)$$

where the curvature parameter $C$ has a constant value of $-0.052$ across the sky and $\nu_0 = 23$ GHz.

### 3.1.3. Model MKD

The MKD model (Martínez-Solaeche et al. 2018) is a realization of thermal dust emission, with the multilayer dust model as re-implemented in a recent version of the Planck Sky Model (Delabrouille et al. 2013; version 2.3.0 of the code). The key idea of the MKD dust model is that if the parameters describing the frequency scaling of the dust emission vary across the sky, they must also vary along the line of sight. The total emission at 353 GHz is modeled as the sum of the emissions from six dust template maps (loosely associated to six layers of distance from the observer), the sum of which is constrained, to give the total dust emission at this frequency. The total emission is obtained from the Planck intensity and polarization data (Planck Collaboration et al. 2016, 2020b), with the GNILC component separation method (Remazeilles et al. 2011a). Explicitly,

$$Q^{MKD}_{\nu,p} = \sum_{l=1}^{6} A^Q_{d,l,p} \left(\frac{\nu}{\nu_0}\right)^{\beta_{d,l,p}} B_\nu(T_{d,l,p})$$
$$U^{MKD}_{\nu,p} = \sum_{l=1}^{6} A^U_{d,l,p} \left(\frac{\nu}{\nu_0}\right)^{\beta_{d,l,p}} B_\nu(T_{d,l,p}), \quad (6)$$

where $l$ identifies a layer of emission. In practice, at high galactic latitude, only the first three layers contribute to the total emission. Closer to the Galactic plane, up to six layers have nonvanishing contributions, with more distant layers contributing the bulk of the emission. The distribution of the dust emission into the various layers is based on measurements of extinction (Green et al. 2015); see Martínez-Solaeche et al. (2018) for more details. We add the "s1" synchrotron to the multilayer dust model to produce the full Galactic emission. While it is common to refer to the multilayer dust of Martínez-Solaeche et al. (2018) as MKD, in this paper the acronyms refer to the full emission, which includes the synchrotron component.

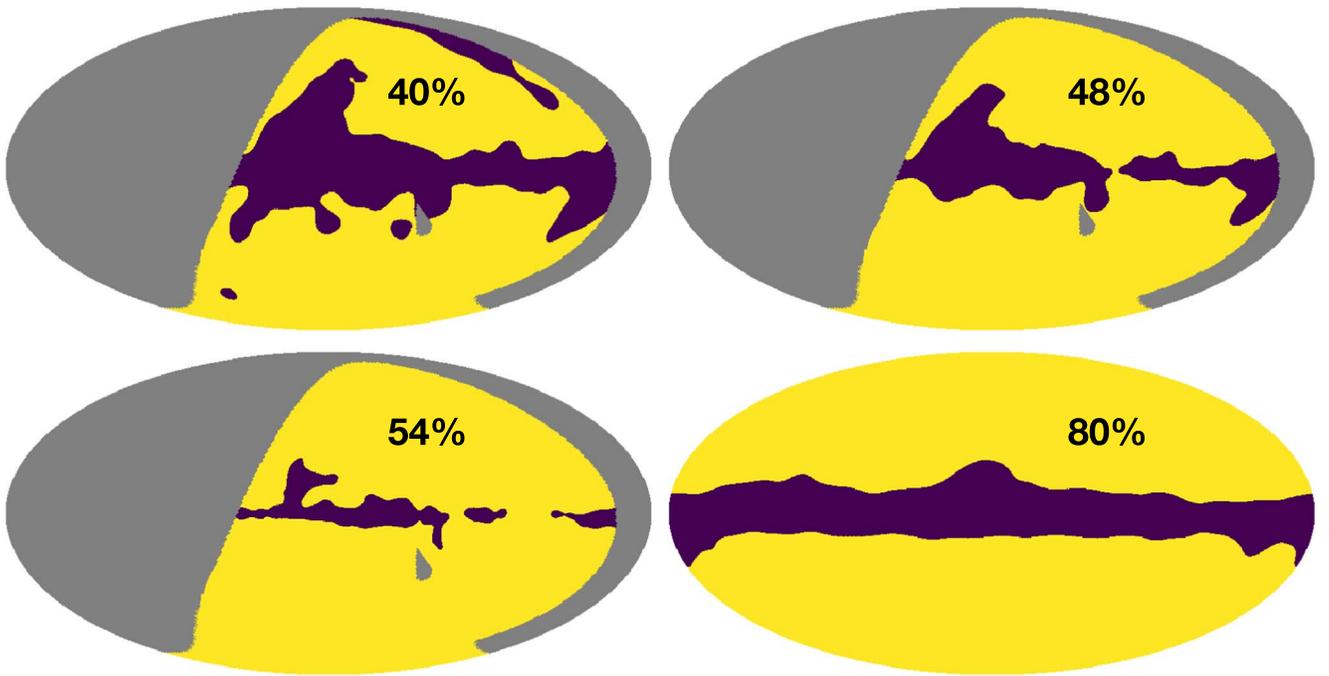

**Figure 2.** Sky masks in Galactic coordinates. The $\tau$S observing region with $f_{sky} = 57\%$ is the combination of yellow and purple regions in the masks, labeled 40%, 48%, and 54%; the gray regions are not observed. In the analysis, we use three masks (purple), removing $f_{sky} = 17\%$, 9%, and 3%, respectively. In Section 5, we also use the Planck mask, labeled 80%, which masks 20% of the sky along the Galactic plane.

### 3.1.4. Model Summary

The Galactic emission models were chosen to encapsulate a range of possible complexity. Model d1s1 is the simplest model that is compatible with current observational constraints, being based on a physically motivated parametric fit of a few parameters that provides an adequate description of the data (Miville-Deschênes et al. 2008; Bennett et al. 2013; Planck Collaboration X 2016). Model d7s3 gives more complexity for the emission components in frequency space, by using a dust model that is not described by an analytic formula, more realistically representing the fact that microwave dust emission arises from grains at a range of temperatures, and a synchrotron model that includes a curvature parameter, as might be expected from the aging of high-energy cosmic-ray electrons. The MKD model expands the line-of-sight complexity of the dust component of d1s1 by integrating the emission in discrete layers. After integration, the resulting SED is no longer described by an MBB emission law, and consequently the spatial variations of the SED are more complex than in d1s1. The true sky likely has each of the complexities introduced by the d7s3 and MKD models, but the level of deviation from simple parametric emission laws is not known. Thus, these models provide physically motivated predictions, but new observational data are needed for definitive determinations.

### 3.2. Sky Maps

We construct sky maps using realizations of noise, CMB signal, and each of the three foreground models discussed in Section 3.1. The CMB signal includes E-mode and lensing B-mode polarization (see, e.g., Lewis & Challinor 2006), $\tau = 0.05$, and other cosmological parameters as given by Planck Collaboration et al. (2020a). There is no inflationary B-mode signal. For a given foreground map, we generate 10 realizations of CMB and noise. An observation of the Stokes parameter $Q$ or $U$ at the frequency $\nu$ and pixel $p$ is given by

$$Q_{\nu,p} = G_\nu \star (Q_{\nu,p}^{fgs} + Q_{\nu,p}^{CMB}) + n_{\nu,p} \quad (7)$$

$$U_{\nu,p} = G_\nu \star (U_{\nu,p}^{fgs} + U_{\nu,p}^{CMB}) + n_{\nu,p}, \quad (8)$$

where $G_\nu \star$ denotes convolution with a circular Gaussian beam, $Q^{fgs}$ and $U^{fgs}$ are the foreground signal contributions, and $n_\nu$ represents white homogeneous noise. The beam size and noise parameters are given in Table 1. We determine the constraints on $\tau$ as a function of the sky fraction $f_{sky}$, using three masks, as shown in Figure 2. The masks have $f_{sky} = 40\%$, 48%, and 54% and mean polarized intensities of 51, 63, and 85 $\mu$K, respectively, based on the Planck 353 GHz map with 2° resolution. Table 2 gives the mean polarized intensity values at the two highest-frequency bands for $\tau$S, 310 and 380 GHz, as a function of mask sky fraction, for the three models of Section 3.1.

## 4. Component Separation

We carry out the foreground separation process with two methods, one using a parametric approach and the other being a foreground-blind approach. In the parametric approach, one assumes that foreground components are characterized by a template map of emission, scaled in frequency using SEDs that have known functional forms and that are characterized by several parameters. Equation (1) is an example of such parameterization. The component separation process consists of finding the foreground template maps, the best-fit SED parameters, and the amplitude of the CMB in each pixel. In this paper, we use the map-based parametric method described by Stompor et al. (2009) and Errard & Stompor (2019), called fgbuster.[11] In blind approaches, one avoids making a priori

---
[11] https://fgbuster.github.io/

Table 2
Mean Dust Polarized Intensity $P = \sqrt{Q^2 + U^2}$ at 310 and 380 GHz with 2° Resolution for the Foreground Models and Sky Masks Used in This Analysis (40%, 48%, and 54%) and the Planck 2015 80% Mask[1]

| | 310 GHz ($\mu$K) | | | | 380 GHz ($\mu$K) | | | |
|---|---|---|---|---|---|---|---|---|
| $f_{\rm sky}$ | 40% | 48% | 54% | 80% | 40% | 48% | 54% | 80% |
| d1s1 | 23 | 29 | 39 | 28 | 66 | 83 | 112 | 81 |
| d7s3 | 24 | 30 | 41 | 29 | 65 | 82 | 109 | 79 |
| MKD | 25 | 32 | 42 | 31 | 73 | 91 | 121 | 88 |

**Note.** The values based on the Planck 353 GHz map with 2° resolution are 51, 63, 85, and 62 $\mu$K, ordered in increasing sky fraction.
[1] http://pla.esac.esa.int/pla

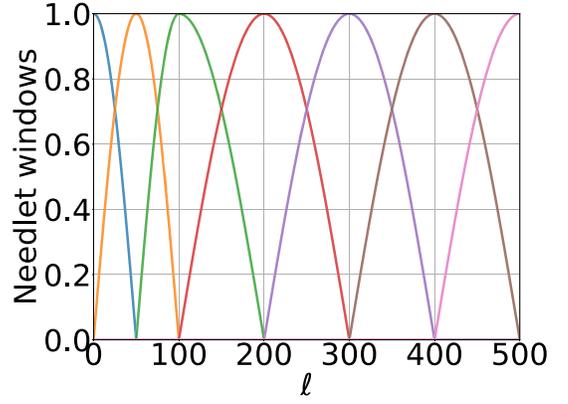

**Figure 3.** Cosine-shaped needlet bandpasses in harmonic space (Basak & Delabrouille 2013). The needlet bandpasses peak at $\ell = 0, 50, 100, 200, 300, 400,$ and 500.

assumptions about the foregrounds. We use the needlet internal linear combination (NILC) technique (Delabrouille et al. 2009), which makes use of the known spectral signature of the CMB and the fact that the CMB is not correlated with foregrounds. No assumptions are made about the number of foreground components, nor their SEDs. For this reason, blind methods are not sensitive to foreground mismodeling (Remazeilles et al. 2016). The NILC method that we adopt here has been used with other data, including WMAP (Basak & Delabrouille 2012, 2013) and Planck (Planck Collaboration et al. 2020b).

We model the data in pixel $p$ as the sum of the CMB, several foreground components, and instrument noise:

$$d_{\nu,p} = a_{\nu 0} c_p + \sum_{k=1}^{N_f} a_{\nu k,p} f_{k,p} + n_{\nu,p}, \quad (9)$$

where $d_{\nu,p}$ is the data at the frequency $\nu$ in pixel $p$, $c_p$ is the CMB signal, and $a_{\nu 0}$ is the frequency scaling of the CMB, which is pixel-independent. The $k$ sum ranges over $N_f$ foregrounds, each described by an amplitude $f_{k,p}$ and an SED $a_{\nu k,p}$. The last term $n_{\nu,p}$ is the noise. This model can be written in vector and matrix format as

$$\boldsymbol{d}_p = \boldsymbol{a} c_p + \boldsymbol{A}_p \boldsymbol{f}_p + \boldsymbol{n}_p, \quad (10)$$

where for each pixel $p$ the data vector $\boldsymbol{d}_p$, the noise $\boldsymbol{n}_p$, and the CMB scaling vector $\boldsymbol{a}$ have as many entries as the number of frequency bands $N_\nu$, the foreground vector $\boldsymbol{f}_p$ has as many entries as the number of distinct foreground emissions $N_f$, and $\boldsymbol{A}_p$ is an $N_\nu \times N_f$ mixing matrix. If one does not single out the CMB, instead treating the CMB and foregrounds on equal footing, Equation (10) can be written as

$$\boldsymbol{d}_p = [\boldsymbol{a} \ \boldsymbol{A}_p] \begin{bmatrix} c_p \\ \boldsymbol{f}_p \end{bmatrix} + \boldsymbol{n}_p, \quad (11)$$

$$\equiv \boldsymbol{\Lambda}_p \boldsymbol{s}_p + \boldsymbol{n}_p, \quad (12)$$

where some of the entries of $\boldsymbol{\Lambda}_p$ can be pixel-dependent. The mixing matrix $\boldsymbol{\Lambda}_p$ now contains both the CMB (the first column of the matrix is $\boldsymbol{a}$) and the foreground scaling laws (the remaining columns of $\boldsymbol{\Lambda}_p$).

For each foreground model, we produce estimates of the CMB E-mode angular power spectrum $\widehat{C}_\ell^{\rm EE}$ with both NILC and fgbuster, for 10 realizations, which have identical foregrounds but different CMB signals and noise. We bin the angular power spectra, which can include residual foregrounds, average them, and calculate the uncertainty for each of the bin powers of width $\Delta \ell = 3$ using (Knox 1995)

$$\widehat{\sigma}_\ell^{\rm EE} = \sqrt{\frac{2}{(2\ell+1) f_{\rm sky} \Delta \ell}} \, \widehat{C}_\ell^{\rm EE}, \quad (13)$$

where $\widehat{C}_\ell^{\rm EE}$ is the average spectrum. We give estimates for $\tau$, having calculated the likelihood (e.g., Tegmark et al. 1997)

$$-2 \ln \mathcal{L}(\tau) = \sum_\ell (2\ell+1) f_{\rm sky} \left[ \frac{\widehat{C}_\ell^{\rm EE}}{C_\ell^{\rm EE, theory}(\tau) + \widehat{N}_\ell} \right.$$
$$\left. + \ln(C_\ell^{\rm EE, theory}(\tau) + \widehat{N}_\ell) \right], \quad (14)$$

where $\widehat{N}_\ell$ is an estimate of the noise power spectrum after component separation. All uncertainty intervals are 68%, unless otherwise noted, and they are derived by integrating the likelihood.

### 4.1. Blind Approach

#### 4.1.1. Detailed Implementation

With NILC, we make no assumption about the foregrounds and include them together with instrumental noise in one single term $\tilde{\boldsymbol{n}}_p$. The data model of Equation (10) becomes:

$$\boldsymbol{d}_p = \boldsymbol{a} c_p + \tilde{\boldsymbol{n}}_p, \quad (15)$$

where the vector $\boldsymbol{a}$ encodes the frequency dependence of the CMB for the frequency bands used and $\tilde{\boldsymbol{n}}_p$ contains all unwanted contributions to the observations.

We decompose the $Q$, $U$ maps using polarized (spin-2 field) spherical harmonic decomposition and obtain the coefficient $a_{\ell,m}^E$ corresponding to the E mode. The NILC method is implemented in a needlet frame (Narcowich et al. 2006). Needlets are a specific type of wavelet, which are localized on the sphere in both the spatial domain and spherical harmonic space. This decomposition makes it possible to optimize the component separation for local variations of both foregrounds and noise as a function of sky location and angular scale. Figure 3 shows the set of harmonic space needlet bandpass windows that are used for this analysis.

For each frequency band, seven maps of E-mode needlet coefficients $d_{\nu,p}^{(j)}$ are obtained by windowing the $a_{\ell m}^E$ coefficients with the seven harmonic space needlet windows $(j) = (1),\ldots,$

(7) of Figure 3 and transforming back to pixel space using an inverse harmonic transform. We estimate maps of the needlet coefficients of the CMB E modes $\hat{s}_p^{(j)}$ at each needlet scale ($j$), by forming a weighted linear combination of the needlet coefficients that minimizes variance under the constraint of preserving the CMB signal

$$\hat{s}_p^{(j)} = \sum_\nu w_{\nu,p}^{(j)} d_{\nu,p}^{(j)}, \qquad (16)$$

where the set of weights $\boldsymbol{w}_p^{(j)} \equiv \{w_{\nu,p}^{(j)}\}$ is given by

$$\boldsymbol{w}_p^{(j)} = \frac{\boldsymbol{a}^{\mathrm{T}}[\boldsymbol{C}_p^{(j)}]^{-1}}{\boldsymbol{a}^{\mathrm{T}}[\boldsymbol{C}_p^{(j)}]^{-1}\boldsymbol{a}}. \qquad (17)$$

The matrix $\boldsymbol{C}_p^{(j)}$ is the data covariance matrix around the pixel $p$, whose elements are estimated for all pairs of frequencies ($\nu, \nu'$) through the convolution

$$C_{p,\nu\nu'}^{(j)} = \int dp' W_{p,p'}^{(j)} d_{\nu,p'}^{(j)} d_{\nu',p'}^{(j)}, \qquad (18)$$

where the Gaussian window functions $W^{(j)}$ have FWHMs of 2.39, 0.93, 0.50, 0.33, 0.27, 0.23, and 0.30 radians for each of the needlet scales, enumerated from low to high $\ell$. We combine the seven maps $\hat{s}_p^{(j)}$ to form an estimate of the final CMB E-mode map. The reconstruction of the CMB E-mode map is performed at a resolution of 33′. All of the steps up to this point, which constitute the NILC component separation steps, are carried out on the entire observed region of the sky that has $f_{\mathrm{sky}} = 57\%$. No masks are applied.

We calculate the E-mode power spectrum $\widehat{C}_\ell^{EE}$ from the reconstructed CMB E-mode map using the MASTER algorithm (Hivon et al. 2002), after applying a chosen Galactic mask. We derive the post–component separation noise power spectrum $\widehat{N}_\ell$ by applying the same needlet decomposition used for the data on noise maps that have frequency band realizations of instrument noise. We calculate the $\tau$ likelihood (Equation (14)) using the signal $\widehat{C}_\ell^{EE}$ and noise $\widehat{N}_\ell$ spectra, using multipoles $2 \leqslant \ell \leqslant 200$. We also calculate and present power spectra of the residual foregrounds and noise. They are obtained by applying the weights $\boldsymbol{w}_p^{(j)}$ to foreground-only maps, to noise-free maps, and to noise-only maps, respectively.

#### 4.1.2. Blind Approach Results

We reconstruct the CMB E-mode power spectra with $\tau$S and $\tau$S-lf for the three foreground input models with sky fraction $f_{\mathrm{sky}} = 40\%$, and we calculate the likelihood for $\tau$. The power spectra results are shown in the left panels of Figure 4. The panels also show spectra of the residual foregrounds, and of the noise and foregrounds combined. The likelihoods of $\tau$ for all the foreground models are shown in the right panels, while central values and 68% confidence intervals are given in Table 3. With two models, d7s3 and MKD, we forecast E-mode reconstruction as a function of sky fraction. The results are given in Table 4.

### 4.2. Parametric Approach

#### 4.2.1. Detailed Implementation

In the parametric approach, one assumes that each of the SEDs has a known functional form that is characterized by a set of parameters $\beta$. The data model of Equation (12) becomes

$$\boldsymbol{d}_p = \boldsymbol{\Lambda}_p(\beta)\,\boldsymbol{s}_p + \boldsymbol{n}_p, \qquad (19)$$

and the component separation process consists of finding the best-fit set of parameters $\tilde{\beta}$ in each pixel, as well as the amplitude of the CMB and the foreground components. We smooth the maps at each frequency band with a Gaussian kernel that has 60′ FWHM, and we carry out the component separation in two steps:

1. maximization of the spectral likelihood, given by (Stompor et al. 2009)

$$-2\log(\mathcal{L})(\beta) = -\sum_p (\hat{\boldsymbol{d}}^T \boldsymbol{N}^{-1}\boldsymbol{\Lambda})(\boldsymbol{\Lambda}^T\boldsymbol{N}^{-1}\boldsymbol{\Lambda})^{-1}(\boldsymbol{\Lambda}^T\boldsymbol{N}^{-1}\hat{\boldsymbol{d}}), \qquad (20)$$

   to produce an estimate of the spectral indices $\tilde{\beta}$, as well as their associated uncertainties. The data $\hat{\boldsymbol{d}}$ denote the smoothed maps and the matrix $\boldsymbol{N}$ is the diagonal noise covariance matrix. We suppress the $p$ indices for clarity; and

2. Estimation of the Q/U amplitudes of the sky signal components:

$$\tilde{\boldsymbol{s}} = (\tilde{\boldsymbol{\Lambda}}^T \boldsymbol{N}^{-1} \tilde{\boldsymbol{\Lambda}})^{-1} \tilde{\boldsymbol{\Lambda}}^T \boldsymbol{N}^{-1} \hat{\boldsymbol{d}}, \qquad (21)$$

   where $\tilde{\boldsymbol{\Lambda}} \equiv \boldsymbol{\Lambda}(\tilde{\beta})$ is the matrix estimated in the first step. We transform the foreground-cleaned estimates of the CMB maps $\tilde{s}^{\mathrm{CMB}}$ into spherical harmonics, produce the bin powers $\widehat{C}_\ell^{EE}$, deconvolve the 60′ smoothing kernel, and calculate the likelihood for $\tau$ as described above.

For the $\tau$S configuration with six frequency bands, $\boldsymbol{\Lambda}$ includes spectral parameters that encode the CMB blackbody emission. Two MBBs – each of the form of Equation (1)—to describe dust emission, and a power law to describe synchrotron emission (see Equation (2)). Of the five parameters, two dust temperatures, two dust spectral indices, and a synchrotron spectral index, we fit only three, fixing one dust temperature to $T_d = 20$ K and the synchrotron spectral index to $\beta_s = -3$, both motivated by data from Planck (Thorne et al. 2017; Planck Collaboration et al. 2020d). Compared to fitting all five spectral indices, this choice reduces the uncertainties on the other derived spectral indices, and therefore also the amplitude of the foreground residuals after component separation, while leaving the bias on $\tau$ small. For all $\tau$S-lf analyses, we fit only the spectral index of one MBB. The other parameters are fixed to $T_d^{(1)} = 30$ K, $\beta_d^{(2)} = 1.4$, $T_d^{(2)} = 20$ K and $\beta_s = -3$. Without a reduction in the parameter space, the component separation process does not converge properly. Other than for $\tau$S-lf, fitting two MBBs with three parameters is the default choice, unless otherwise stated explicitly. In some cases, we compare the derived best-fit signals with their assumed sky models and provide the reduced $\chi^2$ given by

$$\chi^2 = \frac{1}{N_{\mathrm{dof}} - N_{\mathrm{fit}}}(\tilde{\boldsymbol{s}} - \boldsymbol{s})^T \boldsymbol{N}^{-1}(\tilde{\boldsymbol{s}} - \boldsymbol{s}), \qquad (22)$$

where $N_{\mathrm{dof}}$ is the number of degrees of freedom (dof) and $N_{\mathrm{fit}}$ the number of fitted parameters. The matrix operations in Equation (22) are performed over sky pixels, frequencies, and Stokes parameters. We quote the average $\chi^2$ of the 10

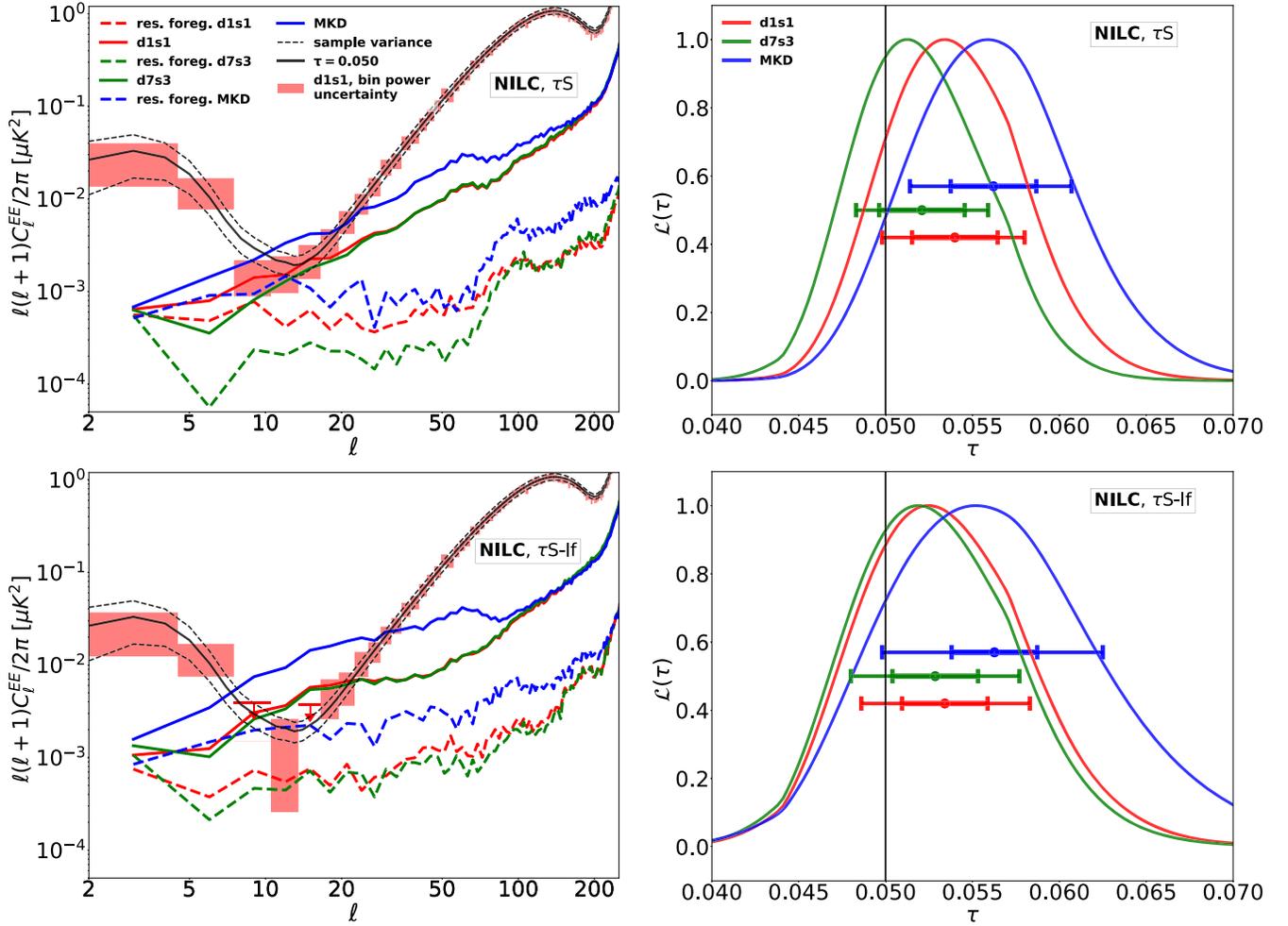

**Figure 4.** Left: NILC E-mode angular power spectra after component separation with $\tau$S (top, red boxes) and $\tau$S-lf (bottom, red boxes; the arrows are the upper limits) for the foreground model d1s1 and the input CMB spectrum with $\tau = 0.05$ (solid black line). Sample variance due to the 40% sky coverage (black dashed line) is a significant source of uncertainty. To maintain clarity, E-mode spectra with the models d7s3 and MKD are not plotted. For all the foreground models, the residual foreground power spectra after component separation (colored dashed lines) are generally small compared to the noise-only spectra (solid colored lines), except at low $\ell$. Right: posterior $\tau$ likelihoods with $\tau$S (top) and $\tau$S-lf (bottom). The error bars give the total 68% interval (outer bars) and that due only to the finite sky fraction (inner bars).

simulations and give the range of probabilities to exceed (PTEs).

As with NILC, the noise $\widehat{N}_\ell$ used in Equation (14) is obtained by applying Equation (21) on noise-only maps, and estimates of the foreground residuals are derived by running the foreground separation on maps that have foregrounds only.

### 4.2.2. Parametric Approach Results

Using fgbuster, we reconstruct the CMB E-mode power spectra for $\tau$S and $\tau$S-lf with the three foreground input models. Bin powers with d1s1, power spectra with all models, and $\tau$ likelihoods are shown in Figure 5, all for $f_{\text{sky}} = 40\%$. The central values for $\tau$ and the 68% confidence intervals are given in Table 3. With two models, d7s3 and MKD, we calculate $\tau$ as a function of sky fraction, and the results are given in Table 4.

## 5. Discussion

### 5.1. Blind Approach

The NILC component separation reconstructs the fiducial value of $\tau = 0.05$ with both $\tau$S and $\tau$S-lf for all foreground models to within $1.4\sigma$; the largest uncertainty value is for $\tau$S-lf with the MKD model. The typical reconstruction is within $1\sigma$. All three peak likelihood $\tau$ values are larger than the input value, indicating that the residual foregrounds contribute to a somewhat biased estimate. Comparing $\tau$S and $\tau$S-lf, we find that the removal of the two higher-frequency bands increases the uncertainty $\sigma(\tau)$ for all models, giving a 20% increase for d1s1, and increasing to 39% for MKD, but there is no increased bias in the $\tau$ estimate.

A significant contribution to the E-mode and $\tau$ uncertainties arises from the limited sky coverage; see Figure 4. We therefore repeat the analysis with smaller Galactic cuts (Figure 2), potentially accepting more residual foreground contamination, but reducing the sample variance by including more sky. An increase of the sky fraction by a factor of 1.35, from 40% to 54%, should have given a 1.16 times decrease in $\sigma(\tau)$ had sample variance been the dominant uncertainty. This analysis is carried out with d7s3 and MKD—the models with the smallest and largest $\sigma(\tau)$ for $f_{\text{sky}} = 40\%$.

With d7s3, we find that the uncertainties decrease by 8%, with no statistically significant increase in bias (see Table 4), indicating that both sky coverage and noise play important

**Table 3**
NILC and fgbuster Forecasts for $\tau$ with Three Foreground Sky Models for $\tau$S and $\tau$S-lf

| | NILC | | |
|---|---|---|---|
| | d1s1 | d7s3 | MKD |
| $\tau$S | $0.0540 \pm 0.0041$ | $0.0521 \pm 0.0038$ | $0.0562 \pm 0.0046$ |
| $\tau$S-lf | $0.0534 \pm 0.0049$ | $0.0529 \pm 0.0049$ | $0.0563 \pm 0.0064$ |

| | fgbuster | | |
|---|---|---|---|
| | d1s1 | d7s3 | MKD |
| $\tau$S[a] | $0.0506 \pm 0.0044$ | $0.0508 \pm 0.0047$ | $0.0515 \pm 0.0050$ |
| $\chi^2_{\tau S}$ | $0.999 \pm 0.004$ | $0.999 \pm 0.004$ | $0.999 \pm 0.004$ |
| min/max PTE | 0.13/0.96 | 0.11/0.96 | 0.12/0.96 |
| $\tau$S-lf[b] | $0.0512 \pm 0.0076$ | $0.0540 \pm 0.0063$ | $0.0525 \pm 0.0089$ |
| $\chi^2_{\tau S-lf}$ | $0.998 \pm 0.006$ | $0.998 \pm 0.006$ | $0.998 \pm 0.006$ |
| min/max PTE | 0.08/0.98 | 0.07/0.97 | 0.08/0.97 |

**Notes.** For fgbuster, we include average reduced $\chi^2$ values, comparing the input and inferred sky signals and the range of PTEs. The analysis uses a mask with $f_{sky} = 40\%$, the input $\tau$ value is 0.05, and the confidence intervals are 68%.
[a] Fitting three spectral parameters.
[b] Fitting one spectral parameter; see the text.

**Table 4**
Forecasts for $\tau$ as a Function of Sky Fraction for the Models d7s3 and MKD and the $\tau$S Experiment Configuration

| | NILC | |
|---|---|---|
| $f_{sky}$ (%) | d7s3 | MKD |
| 40 | $0.0521 \pm 0.0038$ | $0.0562 \pm 0.0046$ |
| 48 | $0.0523 \pm 0.0037$ | $0.0600 \pm 0.0046$ |
| 54 | $0.0527 \pm 0.0035$ | $0.0700 \pm 0.0047$ |

| | fgbuster | |
|---|---|---|
| 40 | $0.0508 \pm 0.0047$ | $0.0515 \pm 0.0050$ |
| 48 | $0.0508 \pm 0.0034$ | $0.0527 \pm 0.0046$ |
| 54 | $0.0513 \pm 0.0034$ | $0.0530 \pm 0.0046$ |

**Note.** The input value is $\tau = 0.05$. The confidence limits are 68% intervals.

roles in determining $\sigma(\tau)$. With MKD, however, a statistically significant bias develops, and there is no decrease in $\sigma(\tau)$. We hypothesize that this is due to residual foreground emission after NILC processing in regions close to the Galactic plane. Table 2 shows increases of ~25% and ~35% in foreground polarized intensity between the 40% and 48% and the 48% and 54% masks, respectively. The increases are consistent with those observed with the Planck 353 GHz map. The MKD model has the highest levels of foreground. As described in Section 4.1, for all sky fractions, the NILC component separation takes place for the entire observed area of the sky, which has $f_{sky} = 57\%$. A mask is only applied when calculating the power spectrum. Bias might not have developed had the component separation been conducted on the partially masked region. Another option for improvement is to use more constrained versions of NILC, which include the deprojection of foreground moments (Remazeilles et al. 2011b, 2021). These refinements are left for future work. With real data, the increase in bias would be identified by applying the analysis as a function of varying sky fractions.

### 5.2. Parametric Approach

The fgbuster component separation reconstructs the fiducial value of $\tau = 0.05$ within $1\sigma$ with both $\tau$S and $\tau$S-lf for all foreground models. The reduced $\chi^2$ values quantifying the difference between the extracted foregrounds and the model foregrounds are consistent with 1, and the PTEs are between 7% and 97%. The largest uncertainty value is for $\tau$S-lf, with the MKD model. As noted in Section 4.2.2, the component separation fits three and only one spectral parameter(s) for $\tau$S and $\tau$S-lf, respectively. With $\tau$S-lf, all three peak likelihood $\tau$ values are larger than the input value, indicating that the residual foregrounds contribute to a somewhat biased estimate. This is also the case with $\tau$S, for the MKD model. Comparing $\tau$S and $\tau$S-lf (see Table 3), we find that the removal of the two higher-frequency bands increases the uncertainty $\sigma(\tau)$ for all models, giving a 25% increase for d7s3 and a near 70% increase for d1s1 and MKD. As with NILC, and perhaps even to a larger degree, the six-band spectral coverage reduces $\sigma(\tau)$. Repeating the analysis for d7s3 and MKD as a function of $f_{sky}$, we find that $\sigma(\tau)$ decreases only between $f_{sky} = 40\%$ and $f_{sky} = 48\%$; see Table 4. With d7s3, the reduction in $\sigma(\tau)$ is larger than expected from the increase in sky coverage, perhaps as a result of small statistics. We observe no statistically significant bias in the value of $\tau$, even with $f_{sky} = 54\%$.

The parametric component separation results that we have presented so far have assumed that the sky can be characterized with two MBBs. We have also assessed the efficacy of component separation as a function of the assumed sky model complexity. For one sky model, d7s3, we repeated the analysis, assuming one and two MBBs with varying numbers of spectral parameters, and with/without fitting for synchrotron emission; see Table 5. The results giving the $\tau$ values at the peak $\tau$ likelihood for $f_{sky} = 54\%$ are shown in Figure 6. All frequency maps are downgraded to a common 60′ resolution. When parameters are not fitted, we fix them at the nominal values used for $\tau$S-lf analyses, $\beta_d^{(2)} = 1.54$, $T_d^{(1)} = 30$ K, $T_d^{(2)} = 20$ K, and $\beta_s = -3$. We find that the component separation with 3 dof with two MBBs minimizes both the statistical error bar on $\tau$ and the bias. Including more dof increases the statistical error bar, whereas decreasing the number of dof increases the bias.

### 5.3. Common Trends and Additional Comments

Although all likelihoods are statistically compatible with the input value, $\tau = 0.050$, the more complex MKD foreground model induces a larger bias and larger uncertainty. Removing the two high-frequency bands, as shown in the bottom panels of Figures 4 and 5, leads to an increase in $\sigma(\tau)$ and could also lead to an increase in bias. This result highlights the benefits of balloon-borne experiments relative to ground-based observations, which are limited to frequencies of less than 300 GHz. Larger sky coverage generally leads to a reduction in $\sigma(\tau)$, without an increase in bias, although at some larger $f_{sky}$ increased bias is likely. We did not investigate this limit.

We consider the values that we derive for $\sigma(\tau)$ to be lower limits, because we assumed white noise that is uniformly distributed across the entire observation area, and because we did not include calibration uncertainties or other systematic effects. Therefore, actual balloon-borne experiments, even if they have an identical configuration to $\tau$S, are not likely to achieve these limits. Ground-based instruments with the same

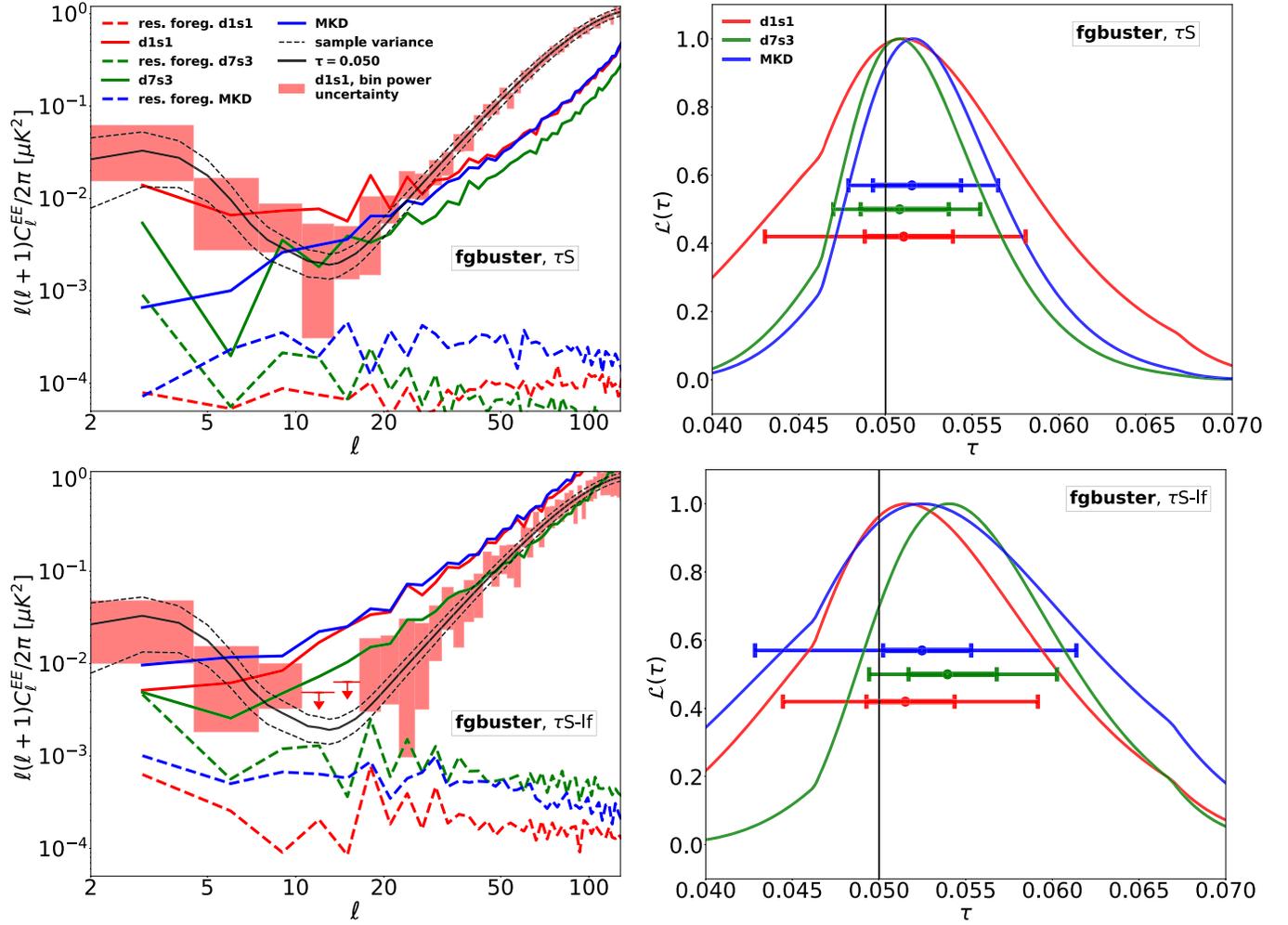

**Figure 5.** The same as Figure 4, but obtained with `fgbuster`.

**Table 5**
Compilation of the Underlying Sky Model Assumptions (Rows), the Numbers of Fitted DoF, and the Variables Being Fit (Columns) Used for Figure 6

|  | $\beta_d^{(1)}$ | $T_d^{(1)}$ | $\beta_d^{(2)}$ | $T_d^{(2)}$ | $\beta_s$ |
|---|---|---|---|---|---|
| 1 MBB/1 dof | ✓ | ⋯ | ⋯ | ⋯ | ⋯ |
| 1 MBB/2 dof | ✓ | ✓ | ⋯ | ⋯ | ⋯ |
| 1 MBB/3 dof | ✓ | ✓ | ⋯ | ⋯ | ✓ |
| 2 MBBs/3 dof | ✓ | ✓ | ✓ | ⋯ | ⋯ |
| 2 MBBs/4 dof | ✓ | ✓ | ✓ | ✓ | ⋯ |
| 2 MBBs/5 dof | ✓ | ✓ | ✓ | ✓ | ✓ |

**Note.** The check marks and dashes indicate the fitted and fixed parameters, respectively. In all cases, the sky model is d7s3.

noise level as $\tau$S are likely to produce even weaker constraints, because of the absence of the high-frequency bands.

At $\tau$S frequency bands and for $\ell < 100$, emission from dust dominates other signals; see Figure 1. However, synchrotron emission, although subdominant, is not negligible, and must be included in the component separation process. Using the three fiducial sky models, which include synchrotron emission, but removing the Planck 30 and 44 GHz band data, which give information about synchrotron emission, gives biased results and larger error bars. With NILC, the removal of these data leads to $\tau$ biases between 2.3$\sigma$ and 3.6$\sigma$, and $\sigma(\tau)$ values that are larger by $\sim$1.5 times compared to the inclusion of the low-frequency data; see the line labeled $\tau$S (no 30/44) in Table 6. With `fgbuster`, the removal of these data gives results that depend on whether we fit only for the CMB and dust (and ignore the existence of a synchrotron component in the underlying sky) or for all three components. When fitting only for the CMB and dust, we find $\tau$ biases between 7$\sigma$ and 17$\sigma$; see the line labeled $\tau$S (dust+cmb) in Table 6. When fitting for the CMB, dust, and a synchrotron component, but still excluding the 30 and 44 GHz data, $\tau$ estimates are unbiased, but $\sigma(\tau)$ values grow by more than a factor of $\sim$5; see the line labeled $\tau$S (no 30/44) in Table 6.

Since the six-band frequency coverage of $\tau$S leads to smaller $\sigma(\tau)$, we examined the impact of having two additional even-higher-frequency bands. We constructed an experiment configuration called $\tau$S-hf, with which we added bands at 450 and 600 GHz. The noise for these bands was scaled up from 380 GHz using the dust spectrum given in Equation (1), with $T_d = 20$ K and $\beta_d = 1.54$ (Thorne et al. 2017; Planck Collaboration et al. 2020d). In contrast with $\tau$S and $\tau$S-lf, we made no effort in this exercise to populate a real focal plane. We repeated the parametric and blind analyses for each of the models, with a fixed $f_{sky} = 40\%$. With NILC, the results, given in Table 6, are essentially the same as with $\tau$S. There are no significant changes with either the central values or with $\sigma(\tau)$.

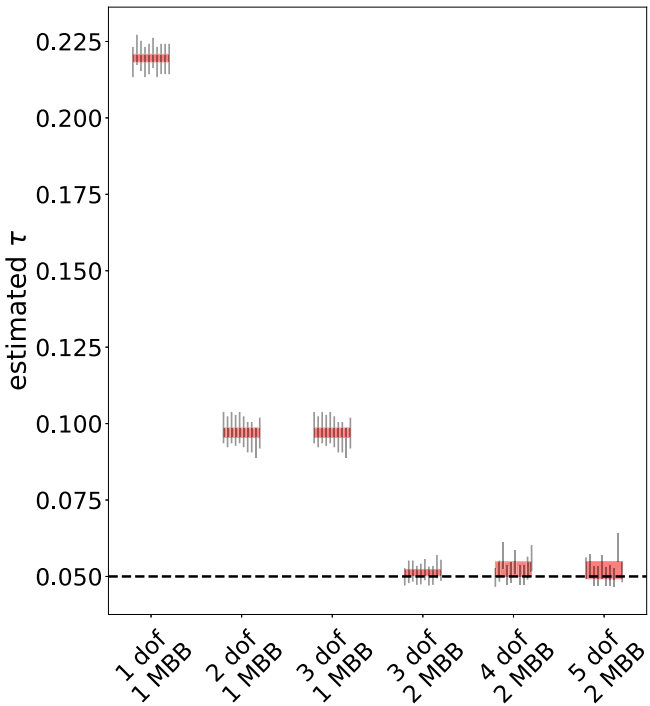

**Figure 6.** fgbuster estimates of $\tau$ and $\sigma(\tau)$ as a function of the assumed sky model complexity. The input sky model is d7s3 and $f_{sky} = 54\%$ in all cases. The parameters used for each of the assumed sky models are listed in Table 5. Each group of points gives the results for 10 CMB + noise simulations. The red band is centered at the average estimate of $\tau$, and its width is the standard deviation of the 10 simulations.

With fgbuster, which used two MBBs and three parameters, there is a significant increase in bias and in $\sigma(\tau)$ for d7s3 and MKD; see Table 6. With MKD, there is nearly $50\sigma$ bias, and the reduced $\chi^2$ clearly indicates a poor fit. All 10 PTEs values for the individual realizations are essentially zero. With d7s3, there is also a large $18\sigma$ bias, but two of the PTE values are 1%. The other eight are between 6% and 36%. This is a case in which foreground separation gives a biased result with acceptable $\chi^2$.

To assess the contribution of instrument noise to the efficacy of each of the foreground separation methods, we repeated the analyses with the $\tau$S configuration, $f_{sky} = 40\%$, and 10 times lower noise—a noise level that is only realistically achievable in space for this sky fraction (Hanany et al. 2019). The results are given in Table 6. Comparing them to Table 3 shows that fgbuster is more sensitive to noise than NILC. With NILC, there is a modest ∼10% reduction in $\sigma(\tau)$, with no significant change in $\tau$. fgbuster gives a noticeable reduction. $\sigma(\tau)$ reaches ∼0.003 with all three models. For MKD, the reduction in $\sigma(\tau)$ is 40%.

A potentially realistic scenario is to deploy a $\tau$S-like experiment twice, in the southern and northern hemispheres, achieving coverage of a larger sky fraction. With a Planck mask that has $f_{sky} = 80\%$[12], and assuming the same noise level as $\tau$S, NILC and fgbuster could give $\sigma(\tau)$ as low as 0.0027 and 0.0024 with d7s3, respectively; see Table 6. We note, however, that with MKD, NILC finds a $3\sigma$ biased value for $\tau$. The value obtained, $\tau = 0.006$, is the same as the one obtained for $f_{sky} = 48\%$ (Table 4), which is consistent with the

---

[12] http://pla.esac.esa.int/pla

---

observation that the levels of the foreground polarized intensities are similar for $f_{sky} = 48\%$ and 80%; see Table 2.

### 5.4. Constraints on $\sum m_\nu$

Improvements in the measurement of $\tau$, in combination with the primary CMB, CMB lensing, and BAO data, are known to give improved constraints on the sum of neutrino masses $\sum m_\nu$ (Lesgourgues & Pastor 2006; Dvorkin et al. 2019). We translate our projections for measurements of $\tau$ into forecasts for $\sigma(\sum m_\nu)$ using standard Fisher matrix techniques. We include BOSS and DESI BAO Fisher matrices, using the same BAO forecasting methods as used by Allison et al. (2015; Font-Ribera et al. 2014). We approximate the current generation of CMB-S3 surveys as having an effective noise of 5 $\mu$K arcmin in the temperature maps ($\sqrt{2}$ larger in polarization) with a 2' beam (Wu et al. 2014). There would be no meaningful change to the results for a 10 $\mu$K arcmin experiment. The CMB-S4 parameters are taken from Abazajian et al. (2019). For both, we include the delensed TT, TE, and EE power spectra, including lensing reconstruction. The lensing reconstruction noise is determined from iterative delensing of both the temperature and polarization maps (Hotinli et al. 2021). We restrict CMB-S3 and CMB-S4 to $\ell > 30$, and include Planck TT data for $\ell < 30$. The BAO and CMB lensing data are assumed to be uncorrelated, which is a good approximation at these noise levels. The $\tau$S uncertainty $\sigma(\tau)$ is included as an external prior, via a Fisher matrix with a single diagonal entry. The Fisher matrix for a given $\sigma(\tau)$ prior is then added to the CMB and BAO Fisher matrices that were computed, assuming the same fiducial value of $\tau = 0.05$.

The results from the combined CMB and BAO data are shown in Figure 7. The vertical lines at $\sigma(\tau) = 0.0034$ and $\sigma(\tau) = 0.005$ represent the smallest and largest values obtained with $\tau$S (not $\tau$S-lf, nor $\tau$S-hf), as indicated in Tables 3 and 4. The smallest and largest $\sigma(\tau)$ values give $\sigma(\sum m_\nu) = 17$, and 20 meV, respectively. The value for 17 meV is lower by only ∼10% compared to $\sigma(\sum m_\nu) = 19$ meV, which is the threshold for $3\sigma$ detection of the minimum sum of neutrino masses $\sum m_\nu = 58$ meV, highlighting the challenge in achieving this threshold. As stated earlier, an experiment would need to achieve an effective noise level of 7 $\mu$K arcmin over 54% of the sky, including statistical and systematic effects, and the foregrounds would need to prove least challenging.

For one foreground model, d7s3, we have shown that a ground-based experiment with a noise level of 7 $\mu$K arcmin and $f_{sky} = 54\%$, but with frequency band coverage only between 150 and 260 GHz, would be limited to $\sigma(\tau) = 0.0041$, which would give a $\sigma(\sum m_\nu)$ constraint tighter than the $3\sigma$ threshold by a few percent; see Figure 7. If the sky is more complex, $\tau$S-lf is not likely to reach the threshold.

The constraints obtained with the CMB-S3 and CMB-S4 data are essentially identical, reflecting the fact that other than $\sigma(\tau)$, the CMB measurements are limited by parameter degeneracies. The DESI BAO data give tighter $\omega_m$ constraints relative to BOSS, due to the more precise BAO measurements, which leads to a factor of ∼1.75 improvement in $\sigma(\sum m_\nu)$.

### 6. Summary

We have assessed the constraints on $\sigma(\tau)$ and $\sigma(\sum m_\nu)$ that are achievable with a next-generation midlatitude balloon-

Table 6
τ Forecasts for Various τS Experiment Configurations

| Name | Short Description | d1s1 | d7s3 | MKD |
|---|---|---|---|---|
| | NILC | | | |
| τS (no 30/44) | Not using 30 and 44 GHz data | $0.0671 \pm 0.0056$ | $0.0620 \pm 0.0052$ | $0.0703 \pm 0.0057$ |
| τS-hf | +450 and 600 GHz | $0.0528 \pm 0.0041$ | $0.0521 \pm 0.0038$ | $0.0558 \pm 0.0046$ |
| τS (noise ×0.1) | Lower noise by ×10 | $0.0534 \pm 0.0037$ | $0.0523 \pm 0.0033$ | $0.0556 \pm 0.0039$ |
| τS-80% | $f_{\rm sky}=80\%$ | $0.0540 \pm 0.0029$ | $0.0520 \pm 0.0027$ | $0.0600 \pm 0.0035$ |
| | fgbuster | | | |
| τS (dust+cmb) | Only fit for dust + CMB | $0.097 \pm 0.0071$ | $0.136 \pm 0.0051$ | $0.115 \pm 0.0060$ |
| τS (no 30/44) | Not using 30 and 44 GHz data | $0.0530 \pm 0.026$ | $0.0518 \pm 0.035$ | $0.0545 \pm 0.050$ |
| τS-hf | +450 and 600 GHz | $0.0510 \pm 0.0034$ | $0.147 \pm 0.0055$ | $0.374 \pm 0.0070$ |
| $\chi^2$ (τS-hf) | | $1.0003 \pm 0.0022$ | $1.0034 \pm 0.0021$ | $29.4 \pm 0.8$ |
| min/max PTE | | 0.09/0.81 | 0.01/0.36 | 0/0 |
| τS (noise ×0.1) | Lower noise by ×10 | $0.0503 \pm 0.0031$ | $0.0508 \pm 0.0031$ | $0.0513 \pm 0.0032$ |
| τS-80% | $f_{\rm sky}=80\%$ | $0.0515 \pm 0.0029$ | $0.0506 \pm 0.0024$ | $0.0550 \pm 0.0037$ |

**Note.** For fgbuster and τS-hf, we include the average $\chi^2$ values, comparing the input and inferred sky signals and the range of PTEs. For all cases except τS-80%, $f_{\rm sky}=40\%$. The confidence limits are 68% intervals.

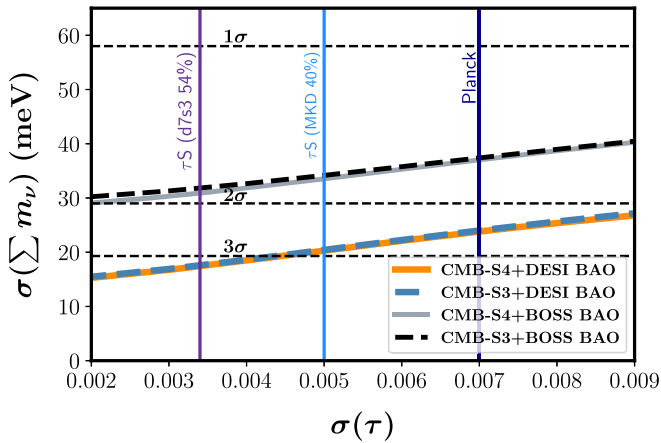

**Figure 7.** Implications of τS constraints on $\sigma(\tau)$ for neutrino mass constraints when combining with CMB-S3/CMB-S4 data and BOSS or DESI BAOs. Two vertical lines bracket the range of $\sigma(\tau)$ achievable with τS and three foreground models, assuming a uniform noise level of 7 μK arcmin over $f_{\rm sky}$, as indicated. The constraints with τS-lf are generally weaker, and in some cases they are mildly stronger with τS-hf. The Planck constraint is from Planck Collaboration et al. (2020a); Pagano et al. (2020) have given a tighter constraint.

borne instrument, and compared these constraints to several other configurations, including one that has no frequency bands above 300 GHz. We find that:

1. The two frequency bands above 300 GHz lead to a smaller $\sigma(\tau)$ relative to a configuration without them. The level of reduction depends on the foreground separation method and foreground model, and is between 20% and nearly 60%;
2. Adding two more frequency bands above 400 GHz does not tighten the constraints on $\tau$;
3. Only under the most optimistic assumptions might both the balloon instrument and its version without the two higher-frequency bands achieve a 3σ detection of the sum of the neutrino masses when the data is combined with small-scale CMB and DESI BAO measurements;
4. Synchrotron emission cannot be neglected. Low-frequency data must be included in the component separation;
5. There is a viable foreground sky model, namely MKD, for which a τS-like instrument will give only a mild improvement, $\sigma(\tau) = 0.005$, beyond the most recent limit of $\sigma(\tau) = 0.006$ (Pagano et al. 2020).
6. Increasing $f_{\rm sky}$ leads to a decrease in $\sigma(\tau)$. While this suggests that teams should opt for larger sky coverage, caution must be taken during the analysis, because statistically significant biases in the estimate of $\tau$ could develop. Analysis that gives constraints on $\tau$ as a function of sky mask could reveal such biases.
7. A more compelling configuration is launching an instrument similar to τS for at least two flights, covering nearly 80% of the sky and potentially giving $\sigma(\tau) < 0.003$, increasing the probability of reaching a 3σ detection of the minimum sum of neutrino masses. However, if the Galactic dust is as the MKD model assumes, the component separation process could give a biased $\tau$ result. As before, analysis that gives constraints on $\tau$ as a function of sky mask could reveal such biases.

We thank an anonymous referee for their useful comments.

**ORCID iDs**

Josquin Errard ● https://orcid.org/0000-0002-1419-0031
Mathieu Remazeilles ● https://orcid.org/0000-0001-9126-6266
Jonathan Aumont ● https://orcid.org/0000-0001-6279-0691
Jacques Delabrouille ● https://orcid.org/0000-0002-7217-4689
Daniel Green ● https://orcid.org/0000-0001-5496-0347
Shaul Hanany ● https://orcid.org/0000-0002-8702-6291
Brandon S. Hensley ● https://orcid.org/0000-0001-7449-4638
Alan Kogut ● https://orcid.org/0000-0001-9835-2351

## References

Abazajian, K. N., & Dodelson, S. 2003, PhRvL, 91, 041301

Abazajian, K., Addison, G., Adshead, P., et al. 2019, arXiv:1907.04473
Abazajian, K. N., Adshead, P., Ahmed, Z., et al. 2016, arXiv:1610.02743
Alam, S., Aubert, M., Avila, S., et al. 2021, PhRvD, 103, 083533
Allison, R., Caucal, P., Calabrese, E., Dunkley, J., & Louis, T. 2015, PhRvD, 92, 123535
Alvarez, M. A., Ferraro, S., Hill, J. C., Hlozek, R., & Ikape, M. 2021, PhRvD, 103, 063518
Amendola, L., Appleby, S., Avgoustidis, A., et al. 2018, LRR, 21, 2
Basak, S., & Delabrouille, J. 2012, MNRAS, 419, 1163
Basak, S., & Delabrouille, J. 2013, MNRAS, 435, 18
Bennett, C. L., Larson, D., Weiland, J. L., et al. 2013, ApJS, 208, 20
Cooray, A., Aguirre, J., Ali-Haimoud, Y., et al. 2019, BAAS, 51, 48
Dahal, S., Amiri, M., Appel, J. W., et al. 2020, JLTP, 199, 289
Dahal, S., Appel, J. W., Datta, R., et al. 2021, ApJ, 926, 33
Datta, R., Austermann, J., Beall, J. A., et al. 2016, JLTP, 184, 568
de Salas, P. F., Forero, D. V., Ternes, C. A., Tortola, M., & Valle, J. W. F. 2018, PhLB, 782, 633
Delabrouille, J., Cardoso, J.-F., Le Jeune, M., et al. 2009, A&A, 493, 835
Delabrouille, J., Betoule, M., Melin, J.-B., et al. 2013, A&A, 553, A96
DESI Collaboration, Aghamousa, A., Aguilar, J., et al. 2016, arXiv:1611.00036
Di Valentino, E., Brinckmann, T., Gerbino, M., et al. 2018, JCAP, 2018, 017
Draine, B. T., & Hensley, B. 2013, ApJ, 765, 159
Dvorkin, C., Gerbino, M., Alonso, D., et al. 2019, BAAS, 51, 64
Errard, J., & Stompor, R. 2019, PhRvD, 99, 043529
Font-Ribera, A., McDonald, P., Mostek, N., et al. 2014, JCAP, 05, 023
Green, D., & Meyers, J. 2021, arxiv:2111.01096
Green, G. M., Schlafly, E. F., Finkbeiner, D. P., et al. 2015, ApJ, 810, 25
Hanany, S., Alvarez, M., Artis, E., et al. 2019, arXiv:1902.10541
Haslam, C. G. T., Salter, C. J., Stoffel, H., & Wilson, W. E. 1982, A&AS, 47, 1
Hensley, B. 2015, PhD thesis, Princeton Univ. http://arks.princeton.edu/ark:/88435/dsp0170795b03v
Hinshaw, G., Nolta, M. R., Bennett, C. L., et al. 2007, ApJS, 170, 288
Hivon, E., Górski, K. M., Netterfield, C. B., et al. 2002, ApJ, 567, 2
Hotinli, S. C., Meyers, J., Trendafilova, C., Green, D., & van Engelen, A. 2021, arXiv:2111.15036
Hu, W., Eisenstein, D. J., & Tegmark, M. 1998, PhRvL, 80, 5255
Hubmayr, J., Appel, J. W., Austermann, J. E., et al. 2011, JLTP, 167, 904
Kaplinghat, M., Knox, L., & Song, Y.-S. 2003, PhRvL, 91, 241301
Knox, L. 1995, PhRvD, 52, 4307
Holland, W. S., Zmuidzinas, J., Lazear, J., et al. 2014, Proc. SPIE, 9153, 91531L
Lesgourgues, J., & Pastor, S. 2006, PhR, 429, 307
Lewis, A., & Challinor, A. 2006, PhR, 429, 1
LiteBIRD Collaboration, Allys, E., Arnold, K., et al. 2022, arXiv:2202.02773
Liu, A., Pritchard, J. R., Allison, R., et al. 2016, PhRvD, 93, 043013
Martínez-Solaeche, G., Karakci, A., & Delabrouille, J. 2018, MNRAS, 476, 1310
Miville-Deschênes, M.-A., Ysard, N., Lavabre, A., et al. 2008, A&A, 490, 1093
Narcowich, F. J., Petrushev, P., & Ward, J. D. 2006, SJMA, 38, 574
Pagano, L., Delouis, J.-M., Mottet, S., Puget, J.-L., & Vibert, L. 2020, A&A, 635, A99
Pan, Z., & Knox, L. 2015, MNRAS, 454, 3200
Pawlyk, S., Ade, P. A. R., Benford, D., et al. 2018, Proc. SPIE, 10708, 1070806
Pelgrims, V., Clark, S. E., Hensley, B. S., et al. 2021, A&A, 647, A16
Planck Collaboration, Aghanim, N., Ashdown, M., et al. 2016, A&A, 596, A109
Planck Collaboration, Aghanim, N., Akrami, Y., et al. 2020a, A&A, 641, A6
Planck Collaboration, Akrami, Y., Ashdown, M., et al. 2020b, A&A, 641, A4
Planck Collaboration, Aghanim, N., Akrami, Y., et al. 2020c, A&A, 641, A1
Planck Collaboration, Akrami, Y., Ashdown, M., et al. 2020d, A&A, 641, A11
Planck Collaboration X 2016, A&A, 594, A10
Planck Collaboration IV 2020, A&A, 641, A4
Remazeilles, M., Delabrouille, J., & Cardoso, J.-F. 2011a, MNRAS, 418, 467
Remazeilles, M., Delabrouille, J., & Cardoso, J.-F. 2011b, MNRAS, 410, 2481
Remazeilles, M., Dickinson, C., Eriksen, H. K. K., & Wehus, I. K. 2016, MNRAS, 458, 2032
Remazeilles, M., Rotti, A., & Chluba, J. 2021, MNRAS, 503, 2478
Stompor, R., Leach, S., Stivoli, F., & Baccigalupi, C. 2009, MNRAS, 392, 216
Tanabashi, M., Hagiwara, K., Hikasa, K., et al. 2018, PhRvD, 98, 030001
Tegmark, M., Taylor, A. N., & Heavens, A. F. 1997, ApJ, 480, 22
The Simons Observatory Collaboration, Ade, P., Aguirre, J., et al. 2018, JCAP, 2019, 56
Thorne, B., Dunkley, J., Alonso, D., & Næss, S. 2017, MNRAS, 469, 2821
Watts, D. J., Wang, B., Ali, A., et al. 2018, ApJ, 863, 121
Wu, W. L. K., Errard, J., Dvorkin, C., et al. 2014, ApJ, 788, 138
Zonca, A., Thorne, B., Krachmalnicoff, N., & Borrill, J. 2021, JOSS, 6, 3783